\documentclass{article}
\usepackage{spconf,amsmath,bm,amssymb}
\usepackage[pdftex]{graphicx}
\usepackage{epstopdf}
\usepackage{algorithm}
\usepackage[noend]{algpseudocode}
\usepackage{pifont}
\usepackage{ifpdf}

\makeatletter
\def\BState{\State\hskip-\ALG@thistlm}
\makeatother

\DeclareMathOperator*{\argmin}{arg\,min}

\title{Large Scale 2D Spectral Compressed Sensing in Continuous Domain}
\name{Jian-Feng Cai$^1$\thanks{JFC is supported in part by Grant 16300616 of Hong Kong Research Grants Council},~~Weiyu Xu$^2$\thanks{Weiyu Xu is supported by the Simons Foundation 318608 , KAUST OCRF-2014-CRG-3, NSF DMS-1418737  and NIH 1R01EB020665-01},~~and Yang Yang$^3$}
\address{\small $^1$Department of Mathematics, Hong Kong University of Science and Technology, Clear Water Bay, Kowloon, Hong Kong SAR, China\\
\small$^2$Department of Electrical and Computer Engineering, University of Iowa, Iowa City, IA 52242 USA\\
\small{$^3$Department of Mathematics, University of Iowa, Iowa City, IA 52242 USA
}}
\numberwithin{equation}{section}
\begin{document}

%
\maketitle
\begin{abstract}
We consider the problem of spectral compressed sensing in continuous domain, which aims to recover a 2-dimensional spectrally sparse signal from partially observed time samples. The signal is assumed to be a superposition of $s$ complex sinusoids. We propose a semidefinite program for the 2D signal recovery problem. Our model is able to handle large scale 2D signals of size 500 $\times$ 500, whereas traditional approaches only handle signals of size around 20 $\times$ 20. 
\end{abstract}
\section{Introduction}
\label{sec:intro}
Spectral compressed sensing aims to recover a spectrally sparse signal from sub-Nyquist sampling of time samples. Consider a frequency-sparse signal $X^\bigstar$  which consists of a superposition of $s$ complex sinusoids with 2-dimensional frequencies $\bm{f}_p = (f_{p1}, f_{p2})$, where $f_{pi}\in [0,1), i=1, 2$. Suppose we observe $m$ regularly-spaced time samples at times $(j,k)$ which form an index set  $T\subset \{0,\ldots,n-1\}\times\{0,\ldots,n-1\}$. We would like to recover the signal of the following form:\footnote{Here we consider square matrices for simplicity, non-square cases can be readily generalized.}
\begin{align}\label{2d}
\begin{aligned}
X^{\bigstar}(j,k)&=\sum_{p=1}^sc_pe^{i2\pi (f_{p1}j+f_{p2}k)}, \\ (j,k)& \in T,
\end{aligned}
\end{align}
where $c_p$ is the coefficient of each complex sinusoid.

In the foundational works of Cand\`es \textit{et al.} \cite{candes2006robust} and Donoho \cite{donoho2006compressed}, successful signal recoveries from very few samples were guaranteed with high probabilities under the assumption that the unknown frequencies lie on a grid. However, in real-world applications, frequencies can possibly lie anywhere on the continuous domain \cite{ekanadham2011recovery, parrish2009improved}. If they are still assumed to lie on a certain grid, the recovery might suffer from the ``basis mismatch'' issue and the recovery might be far from the original true signal \cite{chi2011sensitivity}. Other approaches for frequency-sparse signal recovery have also been proposed, for example, finite rate of innovation (FROI) sampling \cite{vetterli2002sampling, blu2008sparse}. Recently, people have been working on the recovery based on continuous frequency domain. Tang \textit{et al.} \cite{tang2013compressed} showed that $\mathcal{O}(s\log s\log n)$ random samples guarantee exact frequency recovery of a 1D spectrally sparse signal 
\begin{equation}\label{1d signal}
\bm{x}^{\bigstar}(j)=\sum_{p=1}^sc_pe^{i2\pi f_pj}, ~j = 0,\ldots,n-1,
\end{equation} 
with high probability over continuous frequency domain, given the condition that the frequencies are well separated. An atomic norm minimization approach was proposed therein, which was reformulated as an equivalent solvable semidefinite program (SDP). Despite the 1D case, there are more applications involving the recovery of $d$-dimensional signals ($d\geq 2$). For example, super-resolution imaging \cite{rust2006sub} and nuclear magnetic resonance (NMR) spectroscopy \cite{ying2016hankel}.  Chi and Chen \cite{chi2013compressive, chi2015compressive} extended Tang's approach to 2D frequency models. However, the equivalence between the atomic norm minimization and the proposed SDP was not guaranteed. Xu \textit{et al.} \cite{xu2014precise} proposed equivalent SDP formulations of atomic norm minimization to recover signals with 2D off-the-grid frequencies using theories from trigonometric polynomials. Moreover, the SDP formulations were generalized to higher dimensions as well.

In Chi's and Xu's approaches, the $n\times n$ signal matrix $X$ appeared as an $n^2\times 1$ vector in the positive semidefinite constraints. The resulting SDP was on a space of dimension at least $\mathcal{O}(n^2) \times \mathcal{O}(n^2)$. For example, to recover a $100\times 100$ signal $X$, we would have to solve an SDP of size at least $10001\times 10001$, which is computationally prohibitive.

In this paper, we propose a model to recover 2D spectrally sparse signals, which reduces the size of the SDP to $\mathcal{O}(n)\times \mathcal{O}(n)$, while at the same time requires a small number of observed samples to guarantee exact recovery with high probability, which means the model is able to handle large scale 2D signals compared to existing approaches. 

The remainder of this paper is organized as follows. In Section 2, we introduce our proposed SDP formulation for large scale 2D signal recovery. In Section 3, we give numerical experiments to validate our model.

\section{Semidefinite Formulation for 2D Signal Recovery}
Notice that the 2D signal (\ref{2d}) has the matrix form
\begin{equation}\label{form_true}
\begin{aligned}
X^\bigstar & = V_1CV_2^H\\
  & = \sum_{p=1}^sc_p\bm{v_{p1}v_{p2}}^H
\end{aligned}
\end{equation}
where $C$ is a diagonal matrix and $V_1$ and $V_2$ are Vandermonde matrices of the following form{\footnote{In (\ref{form_matrix}), the negative exponents in entries of $V_2$ arise from the conjugate transpose of $\bm{v_{p2}}$ in (\ref{form_true}) in order to be consistent with the original form of the signal (\ref{2d}) .}:

\begin{equation}\label{form_matrix}
\begin{aligned}
C & = \left[
    \begin{array}{ccc}
    c_1 &          &     \\
        & \ddots   &     \\
        &          & c_s
    \end{array}
    \right],\\\\
V_1 & = [\bm{v_{11}}, \cdots, \bm{v_{s1}}]\\ 
    & = \left[
    \begin{array}{ccc}
    e^{i2\pi f_{11}0} & \cdots & e^{i2\pi f_{s1}0}\\
    \vdots           & \ddots & \vdots          \\
    e^{i2\pi f_{11}(n-1)} & \cdots & e^{i2\pi f_{s1}(n-1)}
    \end{array}
    \right],\\\\
V_2 & = [\bm{v_{12}}, \cdots, \bm{v_{s2}}]\\ 
    & = \left[
    \begin{array}{ccc}
    e^{-i2\pi f_{12}0} & \cdots & e^{-i2\pi f_{s2}0}\\
    \vdots           & \ddots & \vdots          \\
    e^{-i2\pi f_{12}(n-1)} & \cdots & e^{-i2\pi f_{s2}(n-1)}
    \end{array}
    \right].
    \\\\    
\end{aligned}
\end{equation}
We assume that $s\ll n$. The $i$th $(i = 1, 2)$ frequency components $f_{pi}$ are distinct across $p$ $(p = 1, \ldots s)$, so that $\text{rank}~(X^\bigstar)=s$ and can be considered low-rank.
The recovery problem can thus be formulated as the following rank minimization problem:
\begin{align}\label{original}
\min_{X,V_1,C,V_2}~~& \text{rank}(X)\nonumber\\
\text{s.t.}~~& 
X=V_1CV_2^H\\
&\text{where}~V_1, V_2~\text{are full rank Vandermonde matrices},\nonumber\\
&C~\text{diagonal as in (\ref{form_matrix})},\nonumber\\
&X(j,k) = X^\bigstar(j,k), (j,k)\in T,\nonumber
\end{align}
where $T$ is the set of indices of the observed entries and $|T| = m$.  

We propose the following semidefinite program for Problem (\ref{original})

\begin{equation}\label{sdp_new}
\begin{aligned}
\min_{T_1,T_2,X}~~& \frac{1}{2}\text{trace}(T_1)+\frac{1}{2}\text{trace}(T_2)\\
\text{s.t.}~~& 
\left[
\begin{array}{cc}
T_1 & X\\
X^H & T_2
\end{array}
\right]\succeq 0,\\
&T_1, T_2~\text{are Toeplitz},\\
&X(j,k) = X^\bigstar(j,k), (j,k)\in T,
\end{aligned}
\end{equation}

The advantage of this SDP over the ones proposed in \cite{chi2015compressive,xu2014precise} is that the size of the semidefinite constraint is $2n\times2n$ instead of $(n^2+1)\times (n^2+1)$, which significantly reduces the memory needed for computation.

\subsection{Why it works}
Our SDP model (\ref{sdp_new}) is inspired from the SDP model in \cite{recht2010guaranteed} 
\begin{equation}\label{sdp}
\begin{aligned}
\min_{W_1,W_2,X}~~& \frac{1}{2}\text{trace}(W_1)+\frac{1}{2}\text{trace}(W_2)\\
\text{s.t.}~~& 
\left[
\begin{array}{cc}
W_1 & X\\
X^H & W_2
\end{array}
\right]\succeq 0,\\
&X(j,k) = X^\bigstar(j,k), (j,k)\in T,
\end{aligned}
\end{equation}
which is equivalent to the problem of minimizing the \emph{nuclear norm} of a matrix $X$: 
\begin{equation}
\begin{aligned}
&\min_{X}~~ ||X||_*\\
\text{s.t.}~~& X(j,k) = X^\bigstar(j,k), (j,k)\in T.
\end{aligned}
\end{equation}
It has been proved to be an effective approach in the \emph{low-rank matrix completion problem} \cite{candes2009exact}.

Our model (\ref{sdp_new}) for 2D signal recovery can be viewed as an extension of the low-rank matrix completion problem in the following sense: 

First, model (\ref{sdp_new}) will yield a low-rank solution. Notice that it minimizes the trace of a positive semidefinite matrix $\left[
\begin{array}{cc}
T_1 & X\\
X^H & T_2
\end{array}
\right]$, which is equivalent to minimizing the sum of all the (nonnegative) eigenvalues. It can be viewed as a relaxation of minimizing the number of nonzero eigenvalues, therefore minimizing rank$\left(\left[
\begin{array}{cc}
T_1 & X\\
X^H & T_2
\end{array}
\right]\right)$ and hence rank$(X)$. 

Second, we would like to incorporate the additional structure of the true signal matrix $X^\bigstar=V_1CV_2^H$ into the recovered solution $X_{\text{rec}}$, therefore imposing the Toeplitz constraints on matrices $T_1$ and $T_2$ so that the problem becomes our model (\ref{sdp_new}). This idea works due to the Caratheodory-Toeplitz Lemma \cite{tang2013compressed}, which states that for Toeplitz positive semidefinite matrices $T_1$ and $T_2$, we have the following decompositions:
\begin{equation}\label{Vander}
T_1 = U_1D_1U_1^H,~~~T_2 = U_2D_2U_2^H
\end{equation}
where $D_1, D_2$ are $r\times r$ diagonal matrices and $U_1, U_2$ are full rank $n\times r$ Vandermonde matrices.
Following the similar argument in \cite{tang2013compressed}, the Vandermonde decomposition (\ref{Vander}) together with the positive semidefinite constraint in (\ref{sdp_new}) imply that $X$ must have the following form
\begin{equation}\label{form_recover}
X = U_1AU_2^H
\end{equation}
Notice that $U_l$ in (\ref{form_recover}) have exactly the same form as $V_l$ in (\ref{form_matrix}) $(l = 1,2)$ (they are all Vandermonde matrices). The only difference between the true signal $X^\bigstar=V_1CV_2^H$ and the recovered signal $X_{\text{rec}}=U_1AU_2^H$ is that the matrix $C$ is required to be diagonal in the true signal, whereas in the recovered signal, there is no such restriction on the matrix $A$, which means that the recovered signal is a structural approximation of the true signal. Therefore, by solving Problem (\ref{sdp_new}), we expect to get a low-rank solution $X_{\text{rec}}$ that resembles the form of the true solution $X^\bigstar$. 

In a nutshell, (\ref{sdp_new}) can be viewed as a low-rank matrix completion problem where the matrix has the additional structure of Vandermonde factors.
We will demonstrate through numerical experiments in Section 3 that the solution $X_{\text{rec}}$ is exactly $X^\bigstar$ with high probability given only a small number of randomly observed time samples.

\section{Numerical Experiments}
We solve (\ref{sdp_new}) by using the alternating direction method of multipliers (ADMM).
To synthesize the $n\times n$ true signal matrix $X^\bigstar$, the $s$ frequency pairs are randomly drawn from $[0,1]^2$ with a minimum separation condition \cite{tang2013compressed} $\Delta_{\text{min}}\triangleq\min_{p\neq q}\{|f_{p1}-f_{q1}|, |f_{p2}-f_{q2}|\}\geq 1/n$, where $p,q = 1, \ldots, s$. The amplitudes $|c_p|$ are drawn randomly from the distribution $0.5+w^2$ with $w$ a zero mean unit variance Gaussian random variable. The phases $e^{i\phi_p}$ are drawn uniformly at random in $[0, 2\pi)$. A total of $m$ observed entries are randomly chosen from $X^\bigstar$. Different values of parameters $m$ and $s$ are used in the implementations of ADMM below to investigate the dependence of $m$ on $s$. The recovery is considered successful if the recovered signal $X_{\text{rec}}$ and the true synthetic signal $X^{\bigstar}$ satisfy $\frac{\|X_{\text{rec}}-X^\bigstar\|_F}{\|X^\bigstar\|_F}\leq 10^{-3}$.

\subsection{ADMM Details}
To apply ADMM, we first reformulate (\ref{sdp_new}) as
\begin{equation}
\begin{aligned}
\min_{M,N}~~& \text{trace}(M)\\
\text{s.t.}~~& M \succeq 0,\\
             & N = 
\left[
\begin{array}{cc}
T_1 & X\\
X^H & T_2
\end{array}
\right],\\
&T_1, T_2~\text{are Toeplitz}\\
&X(j,k) = X^\bigstar(j,k), (j,k)\in T,\\
&M-N = 0.
\end{aligned}
\end{equation}
This new formulation splits the semidefinite constraint and the Toeplitz constraint so that they contain $M$ and $N$ separately.

ADMM consists of the following three updates:
\begin{alignat}{2}
M^{k+1}& =  ~\argmin_M~~&&       \text{trace}(M) + \frac{\rho}{2}\|M-N^k+U^k\|^2_F\\
       &                &&   \text{s.t.} ~~ M \succeq 0, \nonumber\\\nonumber\\
N^{k+1}& =  ~\argmin_N~~&&        \frac{\rho}{2}\|M^{k+1}-N+U^k\|^2_F\\
       &                &&    \text{s.t.} ~~ N = 
\left[
\begin{array}{cc}
T_1 & X\\
X^H & T_2
\end{array}
\right],\nonumber\\
       &   && T_1, T_2~\text{are Toeplitz},\nonumber\\
       &   && X(j,k) = X^\bigstar(j,k), (j,k)\in T,\nonumber\\\nonumber\\
U^{k+1}& = ~U^k+M^{k+1}&&-N^{k+1},\label{u_update}
\end{alignat}
where $\rho$ is a user-specified parameter in ADMM \cite{boyd2011distributed} . We pick $\rho = 0.1$ in the experiments.

M-update can be simplified to
\begin{equation}\label{m_update}
M^{k+1} = VD_{+}V^H,
\end{equation} 
where $D_+$ is $D$ with all negative diagonal entries replaced with 0. And $VDV^H$ is the eigendecomposition of $N^k-U^k-\frac{1}{\rho}I$.

N-update contains 4 blocks that can be simplified to
\begin{align}
&T_1^{k+1}(i+l,i) = \sum_{i=\max\{1,1-l\}}^{\min\{n-l,n\}}[M_1^{k+1}+U_1^{k}](i+l,i)/(n-l),\label{t1_update}\\
&T_2^{k+1}(i+l,i) = \sum_{i=\max\{1,1-l\}}^{\min\{n-l,n\}}[M_2^{k+1}+U_2^{k}](i+l,i)/(n-l),\label{t2_update}\\
&\text{for each~} l = -(n-1), \ldots, 0, \ldots, n-1 \nonumber\\
&\text{and~} \forall i =\max\{1,1-l\}, \ldots, \min\{n-l,n\}.\nonumber\\
&X^{k+1}(j,k) =
\left\{
\begin{array}{ll}
X  ^\bigstar(j,k),                 &\hspace{-0.2cm} (j,k)\in T,\\
\\
\left[M_3^{k+1}+U_3^k\right] (j,k),&\hspace{-0.2cm} \text{otherwise.}
\end{array}     
\right.\label{x_update}
\end{align}

ADMM terminates when both the primal residual $r^k = M^k -N^k$ and dual residual $s^k = \rho(N^{k-1}-N^k)$ satisfy the following conditions, as detailed in \cite{boyd2011distributed}.
\begin{align}
\|r^k\|_F &\leq 2n\epsilon^{\text{abs}}+\epsilon^{\text{rel}}\max\{\|M^k\|_F,\|N^k\|_F\}\label{term1},\\
\|s^k\|_F &\leq 2n\epsilon^{\text{abs}}+\epsilon^{\text{rel}}\|\rho U^k\|_F\label{term2},
\end{align}
where $\epsilon^{\text{abs}}$ and $\epsilon^{\text{rel}}$ are user-specified parameters and here we set both of them to be $10^{-5}$. Once the algorithm terminates, the upper right block of $N$ is the output $X_{\text{rec}}$. The entire ADMM algorithm can be summarized as:
\begin{algorithm}
\caption{ADMM for (\ref{sdp_new})}\label{euclid}
\begin{algorithmic}[1]
\BState $\bm{\textbf{Input}}$ $X^\bigstar$, the observed indices set $T$
\While {(\ref{term1}) and (\ref{term2}) are not satisfied simultaneously\\~~~~~~~~~~}
(\ref{m_update})(\ref{t1_update})(\ref{t2_update})(\ref{x_update})(\ref{u_update})
\EndWhile
\BState $\bm{\textbf{Output}}$ $X_{\text{rec}}$.
\end{algorithmic}
\end{algorithm}

\subsection{Phase Transition}
We investigate the dependence of $m$ on $s$. Here we fix $n = 50$ and implement ADMM with different $m$ and $s$ values. For each $(m,s)$ pair, the simulation is repeated 20 times. The gray level indicates the number of successful recoveries among all 20 repetitions. In Figure 1, we compare between the phase transition plots of the proposed model (\ref{sdp_new}) and the nuclear norm minimization in \cite{recht2010guaranteed} (i.e., after removing the Toeplitz constraint in (\ref{sdp_new})).

\begin{figure}[!htb]
\begin{minipage}[b]{.48\linewidth}
  \centering
  \centerline{\includegraphics[width=4.0cm]{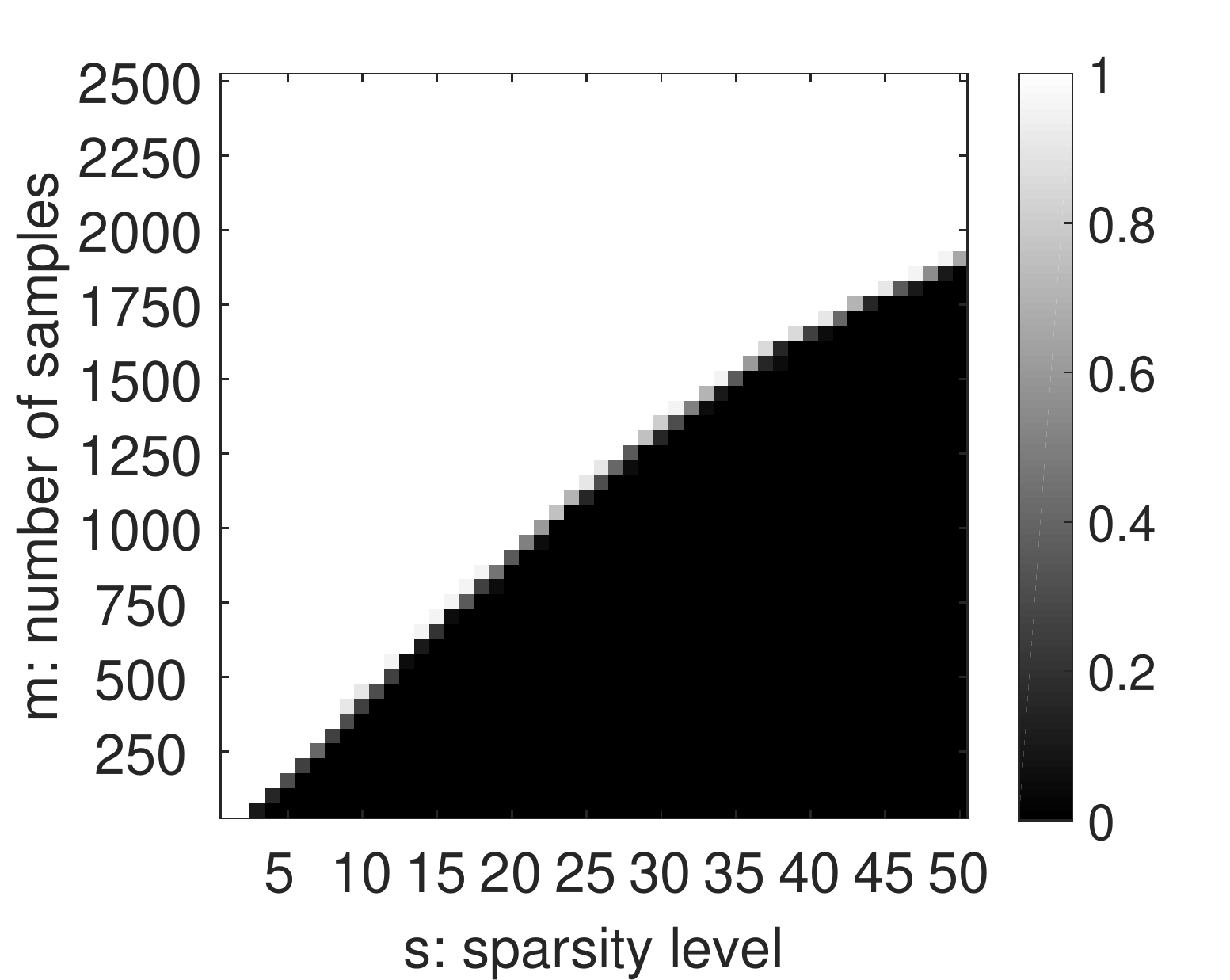}}
\end{minipage}
\hfill
\begin{minipage}[b]{0.48\linewidth}
  \centering
  \centerline{\includegraphics[width=4.0cm]{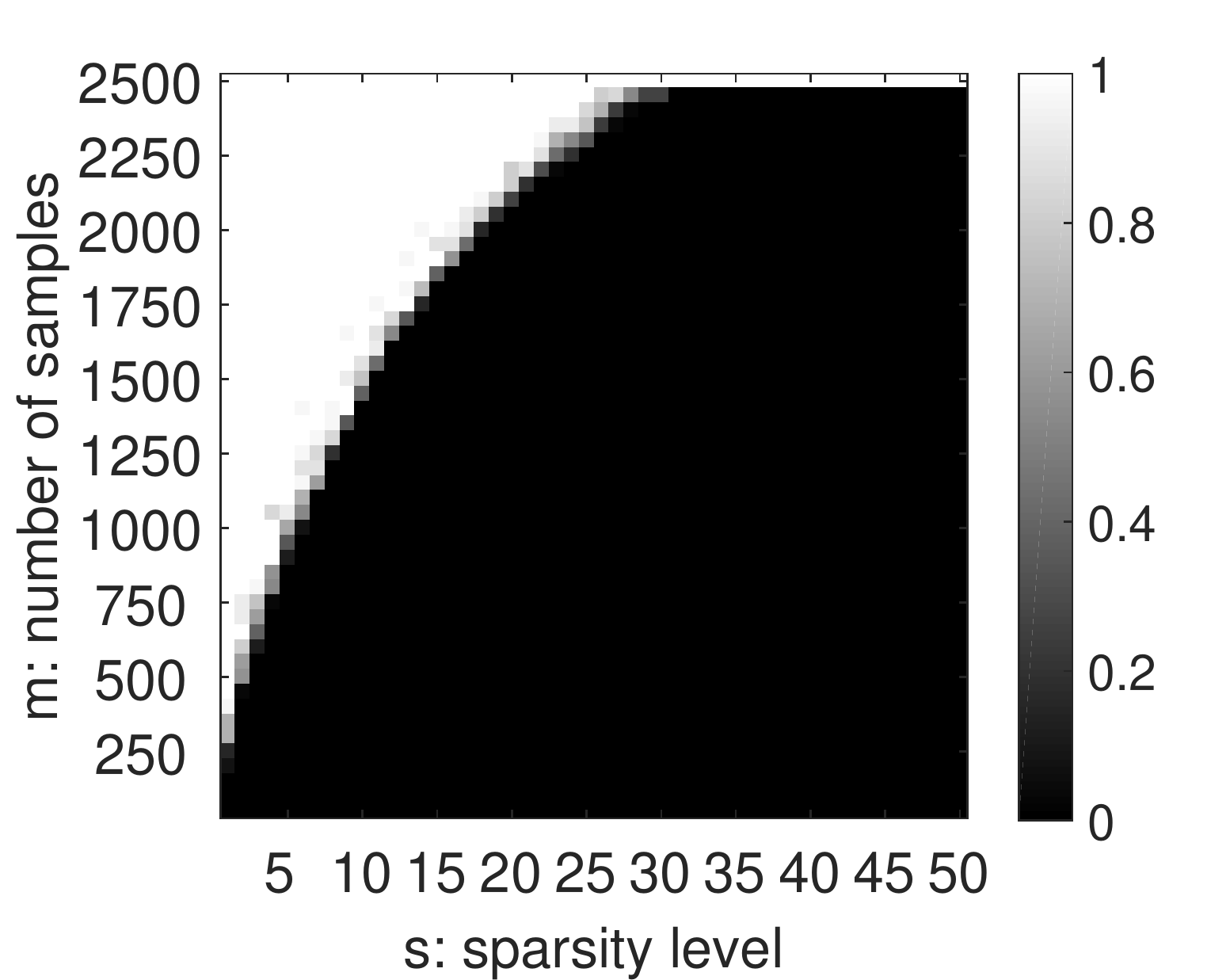}}
\end{minipage}
\caption{$m$ vs $s$ phase transition plots of model (\ref{sdp_new}) (left) and the nuclear norm minimization (right)}
\end{figure}
We can see that the Toeplitz constraint in our model improves the phase transition because it captures the structure of the signal (i.e., the Vandermonde factors). Also, Figure 1 (left) indicates that the dependence of $m$ on $s$ is linear.

\subsection{Comparison with Existing Approach}

Here we compare our approach with Chi's approach (i.e., solving SDP (16) in \cite{chi2015compressive}). We implemented all the experiments on a PC with a 8-core Intel i7-4790 3.60 GHz CPU and 16 GB memory. First, we synthesized 2D signal matrices with different sizes $n$ from $s = 5$ frequency components and randomly drew $m$ time samples. Then we implemented Chi's approach via CVX, and our proposed SDP (\ref{sdp_new}) via both CVX and ADMM. The running times are listed in Table~\ref{tab:a}.
\begin{table}[!htb]
\begin{center}
\scalebox{0.9}{
\begin{tabular*}{1.05\linewidth}{ cccccc }
 \hline\\[-2ex]
 $n$  & $m$ & \cite{chi2015compressive} via CVX & (\ref{sdp_new}) via CVX & (\ref{sdp_new}) via ADMM\\
 \hline              \\[-2ex]   
 15 & 80  & 38.52  & 0.55 & 0.050 \\
 \hline              \\[-2ex]   
 16 & 90  & 60.57  & 0.26 & 0.059\\
 \hline              \\[-2ex]   
 17 & 100  & 103.47 & 0.29 & 0.060\\
 \hline              \\[-2ex]   
 18 & 110  & 161.27 & 0.49 & 0.063\\
 \hline              \\[-2ex]   
 19 & 120  & 239.09 & 0.39 & 0.064\\
 \hline              \\[-2ex]   
 20 & 130 & 363.61 & 0.43 &  0.074\\
 \hline              \\[-2ex]
 21 & 140 & 534.03 & 0.68 &  0.072\\
 \hline              \\[-2ex]
 22 & 150 & 850.19 & 0.54 &  0.076\\
 \hline              \\[-2ex]
 23 & 160 & \text{crashed} & 0.60 & 0.087\\
 \hline
\end{tabular*}
}
\caption{Comparison of running times (in seconds) between existing approach \cite{chi2015compressive} and the proposed SDP formulation (\ref{sdp_new}) for different sizes $n$ and observations $m$}\label{tab:a}
\end{center}
\end{table}
We can see that as the size of the SDP slowly increases, the time required to solve the SDP in \cite{chi2015compressive} grows quickly, whereas the time required for our proposed SDP remains small. Actually, as $n$ goes beyond 23, the SDP in \cite{chi2015compressive} crashes since the size of the SDP is too large to compute. On the contrary, our proposed SDP is still capable of handling large scale problems in a reasonable amount of time, which will be demonstrated in Section \ref{example}.

\subsection{A Large Scale Example}\label{example}
We implemented ADMM on SDP (\ref{sdp_new})  to recover a $500\times 500$ signal generated from $s=10$ random frequencies on $[0,1)\times[0,1)$ with $m=5000$ ($2\%$) randomly observed samples. The ADMM implementation details are demonstrated in Figure ~\ref{fig:large scale}. The relative error $\frac{\|X^\bigstar-X_{\text{rec}}\|_F}{\|X^\bigstar\|_F}=1.4633\times 10^{-4}$, indicating successful recovery. The Fast ADMM by Goldstein \textit{et al.}\cite{goldstein2014fast} was applied to accelerate convergence. The running time is around 24 min on the same PC.
\begin{figure}[!htb]
\begin{minipage}[b]{0.48\linewidth}
  \centering
  \centerline{\includegraphics[width=4.8cm]{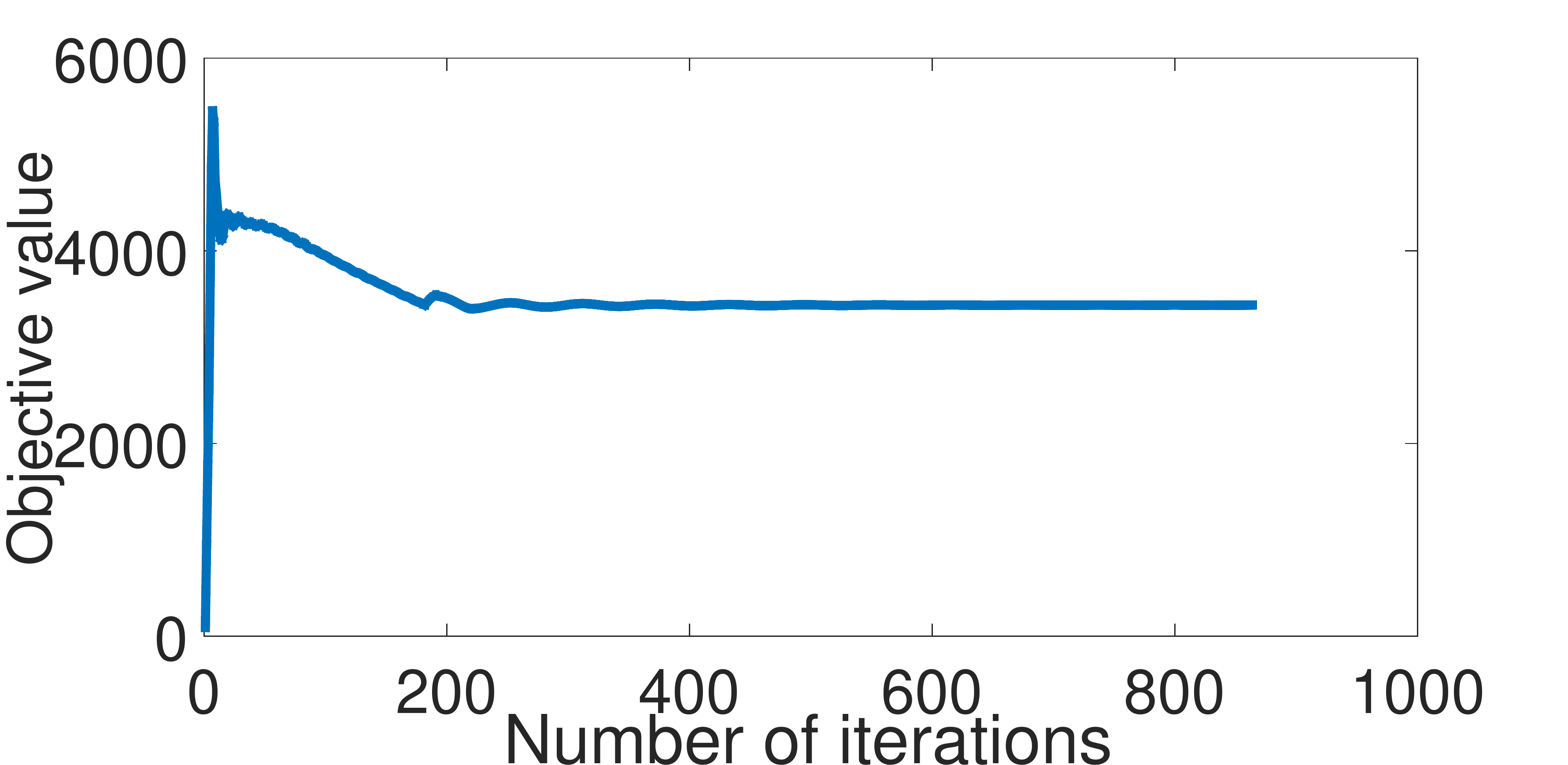}}
\end{minipage}
\hfill
\begin{minipage}[b]{0.48\linewidth}
  \centering
  \centerline{\includegraphics[width=4.7cm]{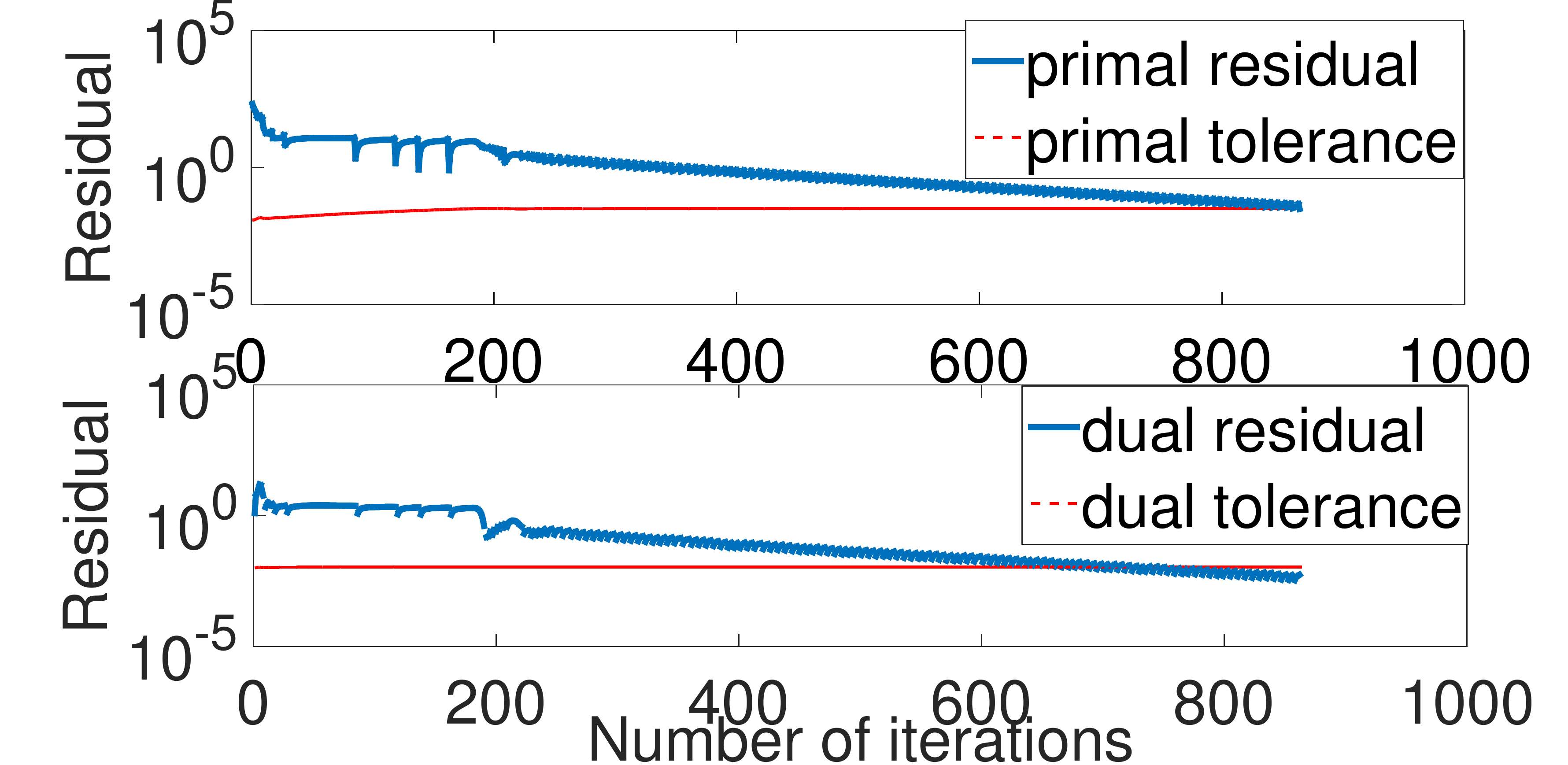}}
\end{minipage}
\caption{Objective values vs number of iterations(left)\newline Primal and dual residuals vs number of iterations (right)}
\label{fig:large scale} 
\end{figure}

The size of SDP (\ref{sdp_new}) for this problem is $1000 \times 1000$. If the approaches in \cite{chi2015compressive, xu2014precise} were adopted, it would become $250,001\times 250,001$, which would be impossible to compute via existing solvers. Therefore, our approach achieved better performance in terms of memory savings.

\section{Conclusion}
In this paper, we proposed a model for the recovery of large scale 2D spectrally sparse signal in continuous domain. This model is able to handle 2D signal matrices with much larger size ($500\times 500$) compared with existing approaches (around $20\times 20$). We demonstrated through numerical experiment that the proposed SDP formulation indeed outperforms existing approaches in terms of running time and memory. 

\newpage
\bibliographystyle{IEEEbib}
\bibliography{strings,refs}

\end{document}